\documentclass[iop,apj]{emulateapj}

\usepackage[USenglish]{babel}
\usepackage{amsmath,amssymb,amsxtra,amsfonts}
\usepackage{graphicx}
\usepackage{hyperref}
\usepackage{natbib}
\newcommand{\lcdm}{\mathrm{\Lambda CDM}}
\newcommand{\densm}{\Omega_{\mathrm{m}}}
\newcommand{\densl}{\Omega_{\mathrm{\Lambda}}}
\newcommand{\densk}{\Omega_{\mathrm{k}}}
\newcommand{\hunit}{\mathrm{km \ s^{-1} \ Mpc^{-1}}}
\newcommand{\likelisym}{\mathcal{L}}
\newcommand{\likelif}[2]{\likelisym\left({#1} | {#2}\right)}
\newcommand{\brset}[1]{\left\{{#1}\right\}}
\newcommand{\der}{\mathrm{d}}
\newcommand{\erf}{\mathrm{erf}}

\shortauthors{Ma \& Zhang}
\shorttitle{Observational Hubble Parameter}
\slugcomment{Draft version December 25, 2010}

\begin{document}

\title{Power of Observational Hubble Parameter Data: a Figure of Merit 
Exploration}

\author{Cong Ma\altaffilmark{1} and Tong-Jie Zhang\altaffilmark{1, 2}}
\altaffiltext{1}{Department of Astronomy, Beijing Normal University, Beijing 
100875, China}
\altaffiltext{2}{Center for High Energy Physics, Peking University, Beijing 
100871, China}
\email{tjzhang@bnu.edu.cn}

\begin{abstract}

We use simulated Hubble parameter data in the redshift range $0 \le z \le 2$ to 
explore the role and power of observational $H(z)$ data in constraining 
cosmological parameters of the $\lcdm$ model.  The error model of the simulated 
data is empirically constructed from available measurements and scales linearly 
as $z$ increases.  By comparing the median figures of merit calculated from 
simulated datasets with that of current type Ia supernova data, we find that as 
many as 64 further independent measurements of $H(z)$ are needed to match the 
parameter constraining power of SNIa.  If the error of $H(z)$ could be lowered 
to $3\%$, the same number of future measurements would be needed, but then the 
redshift coverage would only be required to reach $z = 1$.  We also show that 
accurate measurements of the Hubble constant $H_0$ can be used as priors to 
increase the $H(z)$ data's figure of merit.

\end{abstract}


\section{Introduction}\label{sec:intro}

The expansion of the Universe can be quantitatively studied using the results 
from a variety of cosmological observations, for example the mapping of the 
cosmic microwave background (CMB) anisotropies \citep{2007ApJS..170..377S, 
2010arXiv1001.4538K}, the measurement of baryon acoustic oscillation (BAO) 
peaks \citep{2005ApJ...633..560E, 2010MNRAS.401.2148P}, the linear growth of 
large-scale structures (LSS) \citep{2004PhRvL..92x1302W}, strong gravitational 
lensing \citep{2009MNRAS.394.1449Y}, and measurements of ``standard candles'' 
such as the redshift-distance relationship of type Ia supernovae (SNIa) 
\citep{1998AJ....116.1009R, 2009ApJ...700.1097H} and gamma-ray bursts (GRBs) 
\citep{2004ApJ...613L..13G, 2008ApJ...680...92L}.  Among the various 
observations is the determination of the Hubble parameter $H$ which is directly 
related to the expansion history of the Universe by its definition: $H = 
\dot{a}/{a}$, where $a$ denotes the cosmic scale factor and $\dot{a}$ is its 
rate of change with respect to the cosmic time.  In practice, the Hubble 
parameter is usually measured as a function of the redshift $z$, and the 
redshift is related to $a$ by the formula $a(t) / a(t_0) = 1/(1 + z)$ where 
$t_0$ is the current time that is taken to be a constant.  In the rest of this 
paper we will use the abbreviation ``OHD'' for observational $H(z)$ data.

Though not directly observable, $H(z)$ can nevertheless be deduced from various 
observational data, such as cosmological ages (``standard clocks'') or sizes 
(``standard rods'').  The former method has been illustrated by 
\citet{2002ApJ...573...37J} and leads to OHD found from differential ages of 
galaxies.  The latter method has been discussed by \citet{2003ApJ...594..665B} 
and \citet{2005ApJ...633..575S}, which leads to $H(z)$ found from BAO peaks.  

Moreover, the Hubble parameter as a quantitative measure of the cosmic 
expansion rate is closely related to cosmological distances.  In particular, it 
can be reconstructed from the luminosity distances of SNIa 
\citep{2005PhRvD..71j3513W, 2006MNRAS.366.1081S, 2008A&A...481..295M} and GRBs 
\citep{2010PhRvD..81h3518L}.  On the other hand, given the aforementioned 
availability of {\em independent} determination of $H(z)$, and the expectation 
of more available data in the future, we are interested in the comparison of 
OHD and SNIa data in terms of their respective merits in constraining 
cosmological parameters.  This interest of ours is expressed by the following 
two questions:  Can future observational determinations of the Hubble parameter 
be used as a viable alternative to current SNIa data?  If so, how many more 
datapoints are needed so that the cosmological parameter constraints obtained 
from $H(z)$ data are as good as those obtained from SNIa distance-redshift 
relations?

We attempt to illustrate possible answers to these two questions via an 
exploratory, statistical approach.  This paper is organized as follows: we 
first briefly summarize the current status of available OHD results in Section 
\ref{sec:avail}.  Next, we show how the simulated $H(z)$ datasets are used in 
our exploration in Section \ref{sec:constraint}, and present the results from 
the simulated data in Section \ref{sec:results}.  In Section \ref{sec:future} 
we turn to the data expected from future observation programs.  Finally, in 
Section \ref{sec:conclu} we discuss the limitation and implications of our 
results.

\section{Available Hubble Parameter Datasets}\label{sec:avail}

\begin{figure}
    \includegraphics[width=0.45\textwidth]{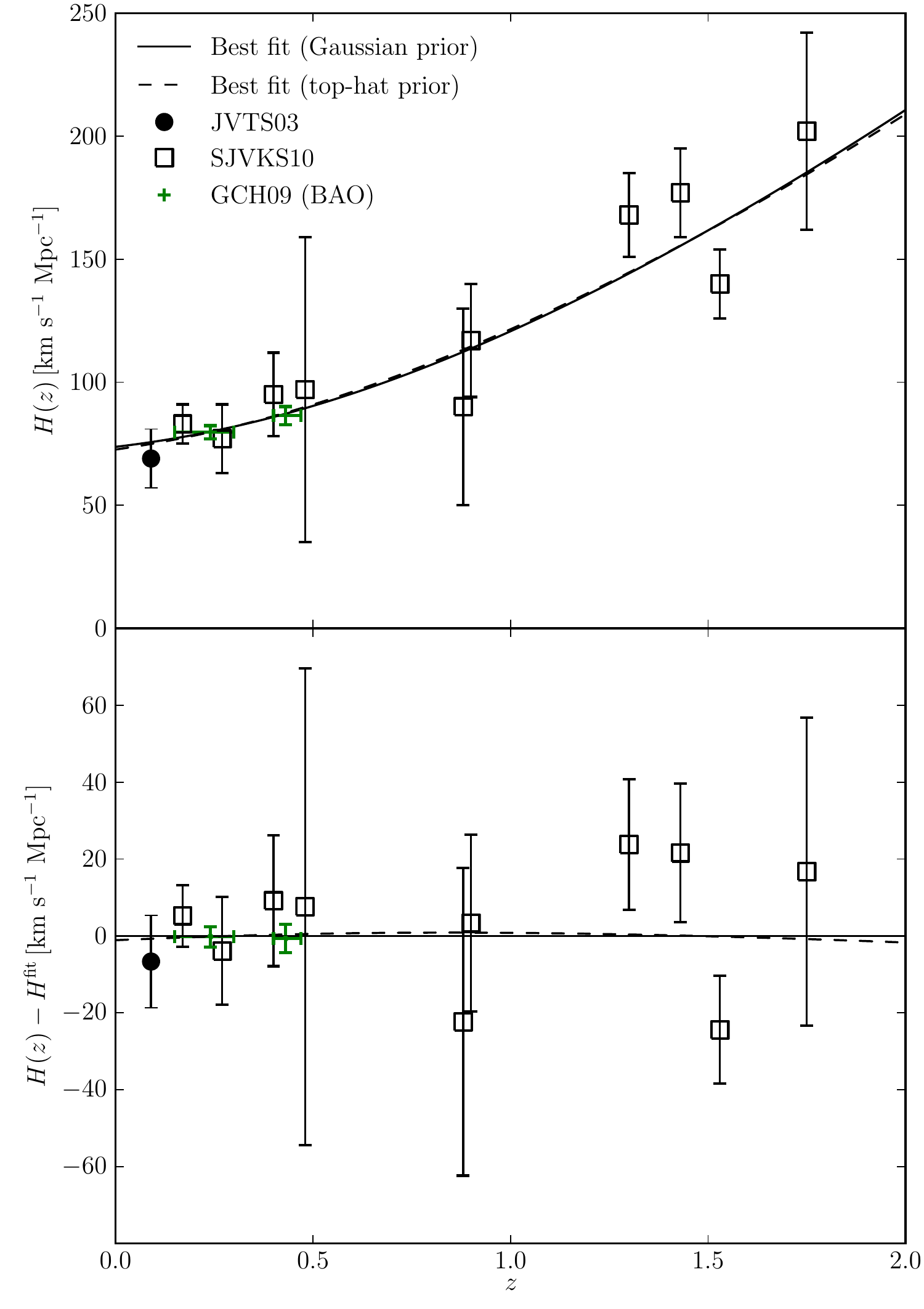}
    \caption{Full OHD set and best-fit $\lcdm$ models.  The top panel shows the 
    dataset and fit results. The bottom panel shows the residuals with respect 
    to the best-fit model with Gaussian prior on $H_0$.  The $H_0$ priors are 
    described in Section \ref{ssec:tmodel}.}
    \label{fig:hzobs}
\end{figure}

Currently the amount of available $H(z)$ data is scarce compared with SNIa 
luminosity distance data.  \citet{2003ApJ...593..622J} first obtained one 
$H(z)$ data point at $z \approx 0.1$ from observations of galaxy ages 
(henceforward the ``JVTS03'' dataset).  \citet{2005PhRvD..71l3001S} further 
obtained 8 additional $H(z)$ values up to $z = 1.75$ from the relative ages of 
passively evolving galaxies (henceforward ``SVJ05'') and used it to constrain 
the redshift dependency of dark energy potential.  \citet{2010JCAP...02..008S} 
obtained an expanded dataset (henceforward ``SJVKS10'') and combined it with 
CMB data to constrain dark energy parameters and the spatial curvature.  
Besides $H(z)$ determinations from galaxy ages, observations of BAO peaks have 
also been used to extract $H(z)$ values at low redshift \citep[henceforward 
``GCH09'']{2009MNRAS.399.1663G}.  These datasets are summarized in Table 
\ref{tab:hzobs} and displayed in Figure \ref{fig:hzobs}.

\begin{deluxetable}{lccc}
    \tablecaption{Currently Available $H(z)$ Datasets\label{tab:hzobs}}
    \tablehead{
	\colhead{$z$} &
	\colhead{SVJ05} &
	\colhead{SJVKS10} &
	\colhead{GCH09\tablenotemark{a}}
    }
    \startdata
    $0.09$\tablenotemark{b} &  $\phn69\pm12\phd\phn$  &  $\phn69\pm12$     &  
    \nodata  \\
    $0.17$  &  $\phn83\pm8.3\phn$     &  $\phn83\pm8\phn$  &  \nodata     \\
    $0.24^{+0.06}_{-0.09}$  &  \nodata  &  \nodata  &  $79.69\pm2.65$     \\
    $0.27$  &  $\phn70\pm14\phd\phn$  &  $\phn77\pm14$     &  \nodata     \\
    $0.4$   &  $\phn87\pm17.4$        &  $\phn95\pm17$     &  \nodata     \\
    $0.43^{+0.04}_{-0.03}$  &  \nodata  &  \nodata  &  $86.45\pm3.68$     \\
    $0.48$  &  \nodata                &  $\phn97\pm62$     &  \nodata     \\
    $0.88$  &  $117\pm23.4$           &  $\phn90\pm40$     &  \nodata     \\
    $0.9$   &  \nodata                &  $117\pm23$        &  \nodata     \\
    $1.3$   &  $168\pm13.4$           &  $168\pm17$        &  \nodata     \\
    $1.43$  &  $177\pm14.2$           &  $177\pm18$        &  \nodata     \\
    $1.53$  &  $140\pm14\phd\phn$     &  $140\pm14$        &  \nodata     \\
    $1.75$  &  $202\pm40.4$           &  $202\pm40$        &  \nodata
    \enddata
    \tablecomments{$H(z)$ figures quoted in this table are in the units of 
    $\hunit$.}
    \tablenotetext{a}{Uncertainties include both statistical and systematic 
    errors: $\sigma^2 = \sigma^2_{\mathrm{sta}} + \sigma^2_{\mathrm{sys}}$.  
    See Section 2.4 and Table 3 of \citet{2009MNRAS.399.1663G}.}
    \tablenotetext{b}{Data in this row are taken from JVTS03.}
\end{deluxetable}

These datasets have seen wide application in cosmological research.  In 
addition to those mentioned above, \citet{2007MPLA...22...41Y} first used the 
SVJ05 dataset to constrain cosmological model parameters.  
\citet{2006ApJ...650L...5S} also used the data to constrain parameters in 
various dark energy models.  Their results are in consistence with other 
observational data, in particular the SNIa.  Besides parameters constraints, 
OHD can also be used as an auxiliary model selection criterion 
\citep{2009JCAP...06..036L}.

In this paper, the OHD sets used are taken from the union of JVTS03, GCH09, and 
SJVKS10.  The SVJ05 dataset, having been replaced by SJVKS10, is no longer 
used, and is listed in Table \ref{tab:hzobs} for reference only.  Using these 
datasets, we find the parameter constraints for a non-flat $\lcdm$ universe 
$\densm = 0.37^{+0.15}_{-0.16}$ and $\densl = 0.93^{+0.25}_{-0.29}$ assuming 
the Gaussian prior $H_0 = 74.2 \pm 3.6\ \hunit$ suggested by 
\citet{2009ApJ...699..539R}.  We also use a conservative, ``top-hat'' prior on 
$H_0$, namely a uniform distribution in the range $[50, 100]$, and obtained 
$\densm = 0.34^{+0.20}_{-0.27}$ and $\densl = 0.86^{+0.44}_{-0.64}$.  These 
$H_0$ priors are discussed in detail in Section \ref{ssec:tmodel}, and their 
effects on the parameter constraints are illustrated in Section 
\ref{sec:results}.

\section{Parameter Constraint with Simulated Datasets}\label{sec:constraint}

In the current absence of more OHD, we turn to simulated $H(z)$ datasets in the 
attempt to explore the answer to the questions raised in Section 
\ref{sec:intro}.  To proceed with our exploration, we must prepare ourselves 
with a) a way of generating simulated $H(z)$ datasets, b) an ``evaluation'' 
model (or class of models) of cosmic expansion in which parameter constraint is 
performed using the simulated data, and c) a quantified measure of the 
datasets' ability of tightening the constraints in the model's parameter space, 
i.e.~a well-defined ``figure of merit'' (FoM).  These topics are discussed in 
detail in the rest of this section.

\subsection{Generation of Simulated Datasets}\label{ssec:gensim}

Our simulated datasets are based on a spatially flat $\lcdm$ model with $\densm 
= 0.27$ and $\densl = 0.73$. This fiducial model is consistent with the 7-year 
Wilkinson Microwave Anisotropy Probe (WMAP) \citep{2010arXiv1001.4538K}, the 
BAO \citep{2010MNRAS.401.2148P}, and SNIa \citep{2009ApJ...700.1097H} 
observations.  Therefore, it summarizes our current knowledge about the recent 
history of cosmic expansion fairly well.  In this fiducial model the Hubble 
parameter is expressible as a function of redshift $z$ by the simple formula
\begin{equation}
    \label{eq:fidlcdm}
    H_{\mathrm{fid}}(z) = H_0 \sqrt{\densm (1 + z)^3 + \densl}
\end{equation}
where $H_0$ is the Hubble constant.

The modelling of the observational data's deviations from the fiducial model, 
as well as the statistical and systematic uncertainties of the data, can be 
rather difficult.  For SNIa, there are planned projects such as the Wide-Field 
Infrared Survey Telescope (WFIRST\footnote{\url{http://wfirst.gsfc.nasa.gov/}}) 
with well-defined redshift distribution of targets \citep{2004astro.ph..5232S} 
and uncertainty model \citep{2004MNRAS.347..909K, 2009NuPhS.194..239H} based on 
which simulated data can be generated.  However, this is not true for OHD, as 
there have not been a formal specification of future observational goals known 
to the authors.  Consequently, we have to work around this difficulty by 
approaching the problem from a phenomenological point of view.

\begin{figure}
    \includegraphics[width=.45\textwidth]{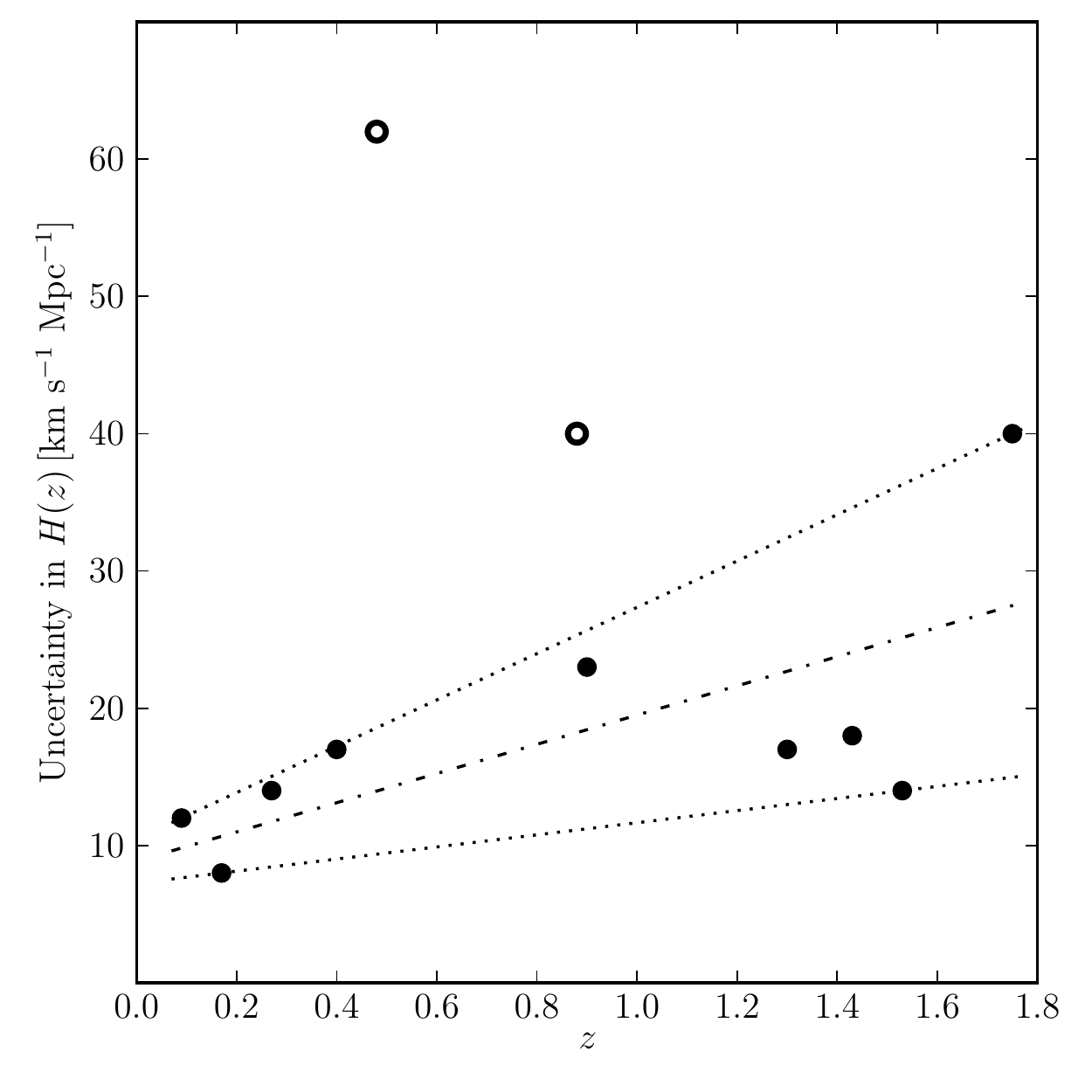}
    \caption{Uncertainties of $H(z)$ in the SJVKS10 dataset.  Solid dots and 
    circles represents non-outliers and outliers respectively.  Our heuristic 
    bounds $\sigma_{+}$ and $\sigma_{-}$ are plotted as the two dotted lines.  
    The dash-dotted line shows our estimated mean uncertainty $\sigma_{0}$.}
    \label{fig:unc}
\end{figure}

By inspecting the uncertainties on $H(z)$ in the SJVKS10 dataset (Figure 
\ref{fig:unc}), we can see that the general trend of the errors' increasing 
with $z$ despite the two outliers at $z = 0.48$ and $0.88$.  Excluding the 
outliers from the dataset, we find that the uncertainties $\sigma(z)$ are 
bounded by two straight lines $\sigma_{+} = 16.87 z + 10.48$ and $\sigma_{-} = 
4.41 z + 7.25$ from above and below respectively.  If we believe that future 
observations would also yield data with uncertainties within the strip bounded 
by $\sigma_{+}$ and $\sigma_{-}$, we can take the midline of the strip 
$\sigma_{0} = 10.64 z + 8.86$ as an estimate of the mean uncertainty of future 
observations.  In our code, this is done by drawing a random number 
$\tilde{\sigma}(z)$ from the Gaussian distribution $N(\sigma_{0}(z), 
\varepsilon(z))$ where $\varepsilon(z) = (\sigma_{+} - \sigma_{-})/4$.  The 
parameter $\varepsilon$ is chosen so that the probability of 
$\tilde{\sigma}(z)$ falling within the strip is $95.4\%$.

Having found a method of generating the random uncertainty $\tilde{\sigma}(z)$ 
for a simulated datapoint, we are able to simulate the deviation from 
$H_{\mathrm{fid}}$.  Namely, we assume that the deviation of the simulated 
observational value from the fiducial, $\Delta H = H_{\mathrm{sim}}(z) - 
H_{\mathrm{fid}}(z)$, satisfies the Gaussian distribution $N(0, 
\tilde{\sigma}(z))$ from which $\Delta H$ can be drawn as a random variable.

Thus a complete procedure of generating a simulated $H(z)$ value at any given 
$z$ is formed.  First, the fiducial value $H_{\mathrm{fid}}(z)$ is calculated 
from equation (\ref{eq:fidlcdm}).  After that, a random uncertainty 
$\tilde{\sigma}(z)$ is drawn using the aforementioned method.  This uncertainty 
is in turn used to draw a random deviation $\Delta H$ from the Gaussian 
distribution $N(0, \tilde{\sigma}(z))$.  The final result of this process is a 
datapoint $H_{\mathrm{sim}}(z) = H_{\mathrm{fid}}(z) + \Delta H$ with 
uncertainty $\tilde{\sigma}(z)$.

In addition to the procedures described above, the Hubble constant (in the 
units of $\hunit$) is also taken as a random variable and is drawn from the 
Gaussian distribution $N(70.4, 1.4)$ suggested by 7-year WMAP results when we 
evaluate
the right-hand side of equation (\ref{eq:fidlcdm}).  We could have fixed $H_0$ 
at a constant value, but as we shall see in Section \ref{ssec:tmodel}, in our 
analysis we treat $H_0$ and $\densm$ quite differently.  Namely, $\densm$ is a 
parameter with a posterior distribution to be inferred, but $H_0$ is a nuisance 
parameter that is marginalized over using some independent measurement results 
as prior knowledge.  This treatment can be found in many works, such as 
\citep{2010JCAP...02..008S} and \citep{2010PhLB..687..286W}.  It can be 
justified by the need to reduce the dimension of the parameter space given 
limited data, and we usually prioritize other parameters such as $\densm$ over 
$H_0$.  Therefore, when generating simulated datasets we sample $H_0$ from a 
random distribution to reflect the uncertainty {\em to be marginalized over}.  
Otherwise, we could have ``cheated'' by forcing a $\delta$-function prior on 
$H_0$ centered at its fiducial value in the analysis of simulated data and 
obtain overly optimistic predictions from the simulated data.

\begin{figure}
    \includegraphics[width=.45\textwidth]{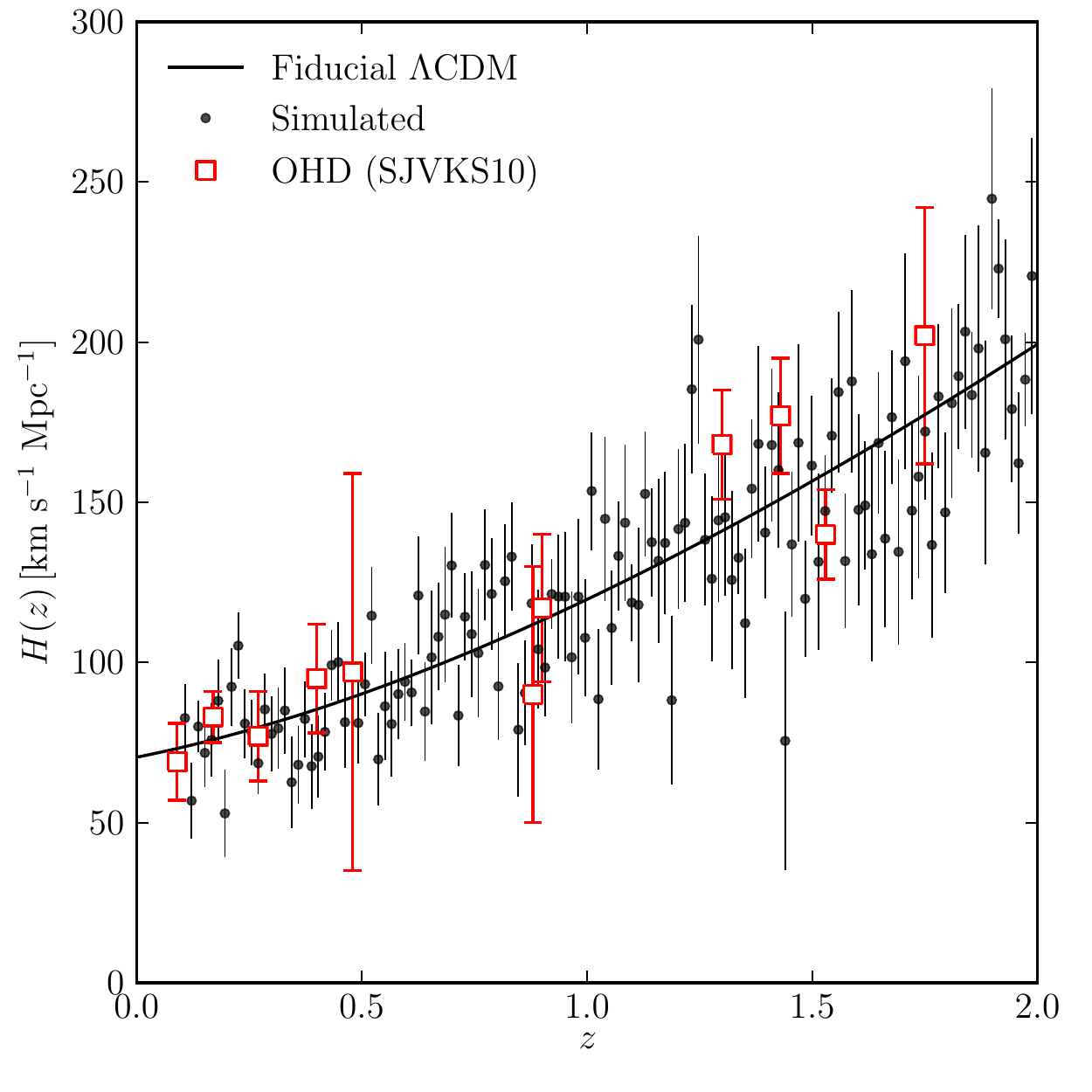}
    \caption{Snapshot of a simulated 
    dataset realized using our method.  
    The uncertainties in the simulated 
    data are modelled 
    phenomenologically after SJVKS10, 
    which is also shown for 
    comparison.}
    \label{fig:snapshot}
\end{figure}

The quality of simulated data thus generated is similar to that of SJVKS10.  A 
snapshot realization of this simulation scheme is displayed in Figure 
\ref{fig:snapshot}, where a total of 128 datapoints with $z$ evenly spaced 
within the range $0.1 \le z \le 2.0$.

\subsection{The Evaluation Model}\label{ssec:tmodel}

We use a standard non-flat $\lcdm$ model with a curvature term $\densk = 1 - 
\densm - \densl$ to evaluate the qualities of the simulated datasets.  In this 
model, the Hubble parameter is given by
\begin{eqnarray}
    \label{eq:testmodel}
    H(z) &=& H_0 \sqrt{\densm (1 + z)^3 + \densk (1 + z)^2 + \densl} \nonumber 
    \\
    &=& H_0 E(z; \densm, \densl).
\end{eqnarray}
Our model choice is mainly motivated by our desire to reduce unnecessary 
distractions arising from the intrinsic complexity of certain cosmological 
models involving dark energy or modified gravitation.

We perform a standard maximal likelihood analysis using this evaluation model.  
In our analysis we intend to marginalize the likelihood function over the 
Hubble parameter $H_0$, thus obtaining parameter constraints in the $(\densm, 
\densl)$ subspace.  This marginalization process also allows us to incorporate 
{\it a priori} knowledge about $H_0$ into our analysis.

There is a fair amount of available information from which reasonable priors 
can be constructed.  \citet{2006ApJ...650L...5S} used two different Gaussian 
priors on $H_0$: one with $H_0 = 73 \pm 3\ \hunit$ from 3-year WMAP data 
\citep{2007ApJS..170..377S}, the other with $H_0 = 68 \pm 4\ \hunit$ by 
\citet{2001ApJ...549....1G} \citep[for a discussion of the non-Gaussianity of 
the error distribution in $H_0$ measurements, see][]{2003PASP..115.1269C}.  
\citet{2009MPLA...24.1699L} used $H_0 = 72 \pm 8\ \hunit$ as suggested by 
\citep{2001ApJ...553...47F}.  In our work we use a more recent determination of 
$H_0 = 74.2 \pm 3.6\ \hunit$ \citep{2009ApJ...699..539R} as an update to the 
ones cited above.  We also consider a ``top-hat'' prior, i.e.~a uniform 
distribution in the interval $[50, 100]$.  Compared with any of the peaked 
priors above, this prior shows less preference of a particular central value 
while still characterizing our belief that any value of $H_0$ outside the said 
range is unlikely to be true.  Notice that the intrinsic spread of $H_0$ 
involved in the generation of simulated datasets is smaller than either prior 
chosen in this step, in consistent with the belief that the prior adopted in 
the estimation of parameters should not be spuriously optimistic.

Having chosen the priors on $H_0$, it is straightforward to derive, up to a 
non-essential multiplicative constant, the posterior probability density 
function (PDF) of parameters given the dataset $\brset{H_i}$ by means of Bayes' 
theorem:
\begin{eqnarray}
    \label{eq:posterior}
    P(\densm, \densl | \brset{H_i}) = \int P(\densm, \densl, H_0 | \brset{H_i}) 
    \ \der H_0 \nonumber \\
   = \int \likelif{\brset{H_i}}{\densm, \densl, H_0} P(H_0) \ \der H_0 \\
   \nonumber
\end{eqnarray}
where $\likelisym$ is the likelihood and $P(H_0)$ is the prior PDF of $H_0$.  
Assuming that each measurement in $\brset{H_i}$ has Gaussian error distribution 
of $H(z)$ and is independent from other measurements, the likelihood is then 
given by
\begin{equation}
    \label{eq:likeli}
    \likelif{\brset{H_i}}{\densm, \densl, H_0} = \left(\prod_{i} 
    \frac{1}{\sqrt{2\pi\sigma^{2}_{i}}}\right) \exp \left( -\frac{\chi^2}{2} 
    \right),
\end{equation}
where the $\chi^2$ statistic is defined by
\begin{equation}
    \label{eq:chisq}
    \chi^2 = \sum_i \frac{\left[H_{0}E(z_i; \densm, \densl) - 
    H_i\right]^2}{\sigma^2_i},
\end{equation}
and $\sigma_i$'s are the uncertainties quoted from the dataset.  The posterior 
PDF can thus be found by inserting equation (\ref{eq:likeli}) and the exact 
form of $P(H_0)$ into equation (\ref{eq:posterior}).

We now show that the integral over $H_0$ in equation (\ref{eq:posterior}) can 
be worked out analytically for our $P(H_0)$ choices.

\paragraph{Gaussian prior.}

Let
\begin{equation}
    \label{eq:gaussianprior}
    P(H_0) = \frac{1}{\sqrt{2 \pi \sigma^2_H}} \exp \left[ -\frac{\left(H_0 - 
    \mu_H\right)^2}{2\sigma^2_H} \right].
\end{equation}

In this case, equation (\ref{eq:posterior}) reduces to
\begin{eqnarray}
    \label{eq:postfromgaussian}
    P(\densm,&& \densl | \brset{H_i}) \nonumber \\
    &&= \frac{1}{\sqrt{A}} \left[ \erf\left( \frac{B}{\sqrt{A}} \right) + 1 
    \right] \exp\left( \frac{B^2}{A} \right)
\end{eqnarray}
where
\begin{eqnarray}
    A &=& \frac{1}{2\sigma^2_H} + \sum_i \frac{E^2(z_i; \densm, 
    \densl)}{2\sigma^2_i}, \nonumber \\
    B &=& \frac{\mu_H}{2\sigma^2_H} + \sum_i \frac{E(z_i; \densm, \densl) 
    H_i}{2\sigma^2_i}, \nonumber
\end{eqnarray}
and $\erf$ stands for the error function.  The form shown in equation 
(\ref{eq:postfromgaussian}) is not normalized; all multiplicative constants 
have been discarded.

\paragraph{Top-hat prior.}

The PDF of a uniform distribution over the interval $[x, y]$ can be written as 
$P(H_0) = \Theta(H_0 - x) \Theta(y - H_0) / (y - x)$ where $\Theta$ denotes the 
Heaviside unit step function.  With this prior on $H_0$, equation 
(\ref{eq:posterior}) becomes
\begin{eqnarray}
    \label{eq:postfromtophat}
    P(\densm,&& \densl | \brset{H_i}) \nonumber \\
    &&= \frac{U(x, C, D) - U(y, C, D)}{\sqrt{C}} \exp\left( \frac{D^2}{C} 
    \right)
\end{eqnarray}
where
\begin{eqnarray}
    C = \sum_i \frac{E^2(z_i; \densm, \densl)}{2\sigma^2_i}, \quad
    D = \sum_i \frac{E(z_i; \densm, \densl) H_i}{2\sigma^2_i}, \nonumber
\end{eqnarray}
and
\begin{eqnarray}
    U(x, \alpha, \beta) = \erf\left(\frac{\beta - x 
    \alpha}{\sqrt{\alpha}}\right).  \nonumber
\end{eqnarray}
The normalization constant has been dropped from formula 
(\ref{eq:postfromtophat}) as well.

\subsection{Figure of Merit}\label{ssec:fom}

The posterior probability density functions obtained above put statistical 
constraints over the parameters.  As the dataset $\brset{H_i}$ improves in size 
and quality, the constraints are tightened.  To evaluate a dataset's ability of 
tightening the constrains, a quantified figure of merit (FoM) must be 
established.

We note that the FoM can be defined arbitrarily as long as it reasonably 
rewards a tight fit while punishing a loose one.  Its definition can be 
motivated purely statistically, for example the reciprocal hypervolume of the 
$95\%$ confidence region in the parameter space \citep{2006astro.ph..9591A}.  
However, a definition that is sensitive to some physically significant 
structuring of the parameter space can be preferable if our scientific goal 
requires it \citep{2006APh....26..102L}.

In this paper we use a statistical FoM definition similar to the ones of 
\citet{2006astro.ph..9591A}, \citet{2008MNRAS.388..275L}, and 
\citet{2009JCAP...11..029B}. Our FoM is defined to be the area enclosed by the 
contour of $P(\densm, \densl | \brset{H_i}) = 
\exp(-\Delta\chi^2/2)P_{\mathrm{max}}$ where $P_{\mathrm{max}}$ is the maximum 
value of the posterior PDF, and the constant $\Delta\chi^2$ is taken to be 
$6.17$.  This value is so chosen that the region enclosed by this contour 
coincides with the $2\sigma$ or $95.4\%$ confidence region if the posterior is 
Gaussian.  In the rest of this paper we will use the term ``$2\sigma$ region'' 
to refer to the region defined in this way.  The $1\sigma$ and $3\sigma$ 
regions can be defined in the same manner by setting the constant 
$\Delta\chi^2$ to $2.3$ and $11.8$ respectively.  It is important to bear in 
mind that our definition of ``$n \sigma$ regions'' are motivated by 
nomenclature brevity rather than mathematical concreteness, since the posterior 
PDFs (eqs.~[\ref{eq:postfromgaussian}] and [\ref{eq:postfromtophat}]) are 
manifestly non-Gaussian.

We make two further remarks on our FoM definition. First, since our definition 
of FoM is statistical rather than physical, we do not exclude the unrealistic 
parts of the confidence regions from the area.  Second, the FoM by our 
definition is obviously invariant under the multiplication of $P(\densm, \densl 
| \brset{H_i})$ by a positive constant, which justifies the omission of 
normalization constants in Section \ref{ssec:tmodel}.

\section{Results from the Simulated Datasets}\label{sec:results}

Using the method described in Section \ref{ssec:gensim}, we generate 500 
realizations of the simulated $H(z)$ dataset.  Each realization contains 128 
datapoints evenly spaced in the redshift range $0.1 \le z \le 2.0$.  From each 
realization, successively shrinking subsamples of 64, 32, and 16 datapoints are 
randomly drawn.  These subsamples are then used in conjunction with the full 
OHD set to obtain their respective figures of merit.

The figures of merit are naturally divided into 4 groups by the size of the 
simulated subsample used in the calculation.  For each group, median and median 
absolute deviation (MAD) statistics are calculated.  The median and the MAD are 
used to represent the central value and spread of FoM data respectively, and 
they are used in preference to the customary pair of the mean and the standard 
deviation, because they are less affected by egregious outliers 
\citep[see][Chapter 1.3.5.6]{nistsematech}.  The outliers mainly arise from 
``worst case'' realizations of simulated $H(z)$ data, for example one with a 
large amount of datapoints deviating too much (or too little) from the fiducial 
model.

To compare the parameter constraining abilities of our simulated $H(z)$ 
datasets with those of SNIa data, we have also fitted our evaluation model 
(eq.~[\ref{eq:testmodel}]) to the ConstitutionT dataset which is a subset of 
the Constitution redshift-distance dataset \citep{2009ApJ...700.1097H} deprived 
of outliers that account for internal tensions
\citep{2010PhLB..687..286W}.  It is worthwhile to point out that the prior of 
$H_0$ used in the SNIa fitting procedure is fundamentally different from either 
one discussed in Section \ref{ssec:tmodel}.  Namely, when SNIa data is used, 
the parameter $H_0$ and the intrinsic absolute magnitude of SNIa, $M_0$, are 
combined into one ``nuisance parameter'' $M = M_0 - 5 \lg H_0$, and is 
marginalized over under the assumption of a flat prior over $(-\infty, 
+\infty)$ \citep{1999ApJ...517..565P}.  This discrepancy should be kept in mind 
when comparing or combining SNIa and $H(z)$ datasets, and we hope it could be 
closed in the future by better constraints, either theoretical or 
observational, on $M_0$.

\begin{figure}
    \includegraphics[width=.45\textwidth]{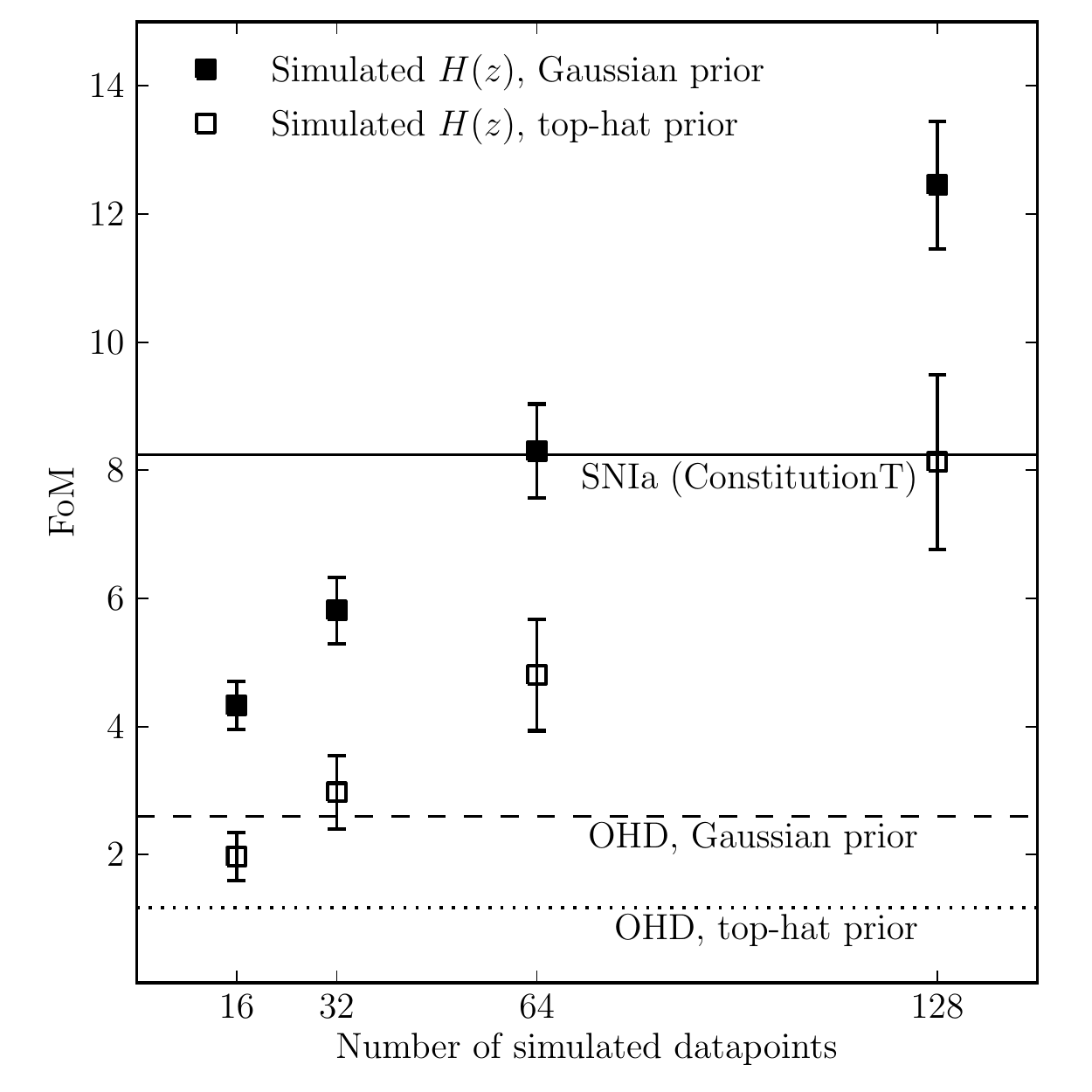}
    \caption{Figures of merit from each group found by combining all 500 
    realizations.  FoM medians are plotted against the sizes of the simulated 
    subsample, and median absolute deviations are shown as error bars.  
    Horizontal lines across the figure marks the figures of merit of purely 
    observational datasets.}
    \label{fig:fom}
\end{figure}

Our main results are show in Figure \ref{fig:fom}.  As one may intuitively 
assume, the median FoM increases with the size of the dataset.  We find that 
the data subset with 64 simulated datapoints leads to a FoM of $8.6\pm0.7$ 
under the Gaussian prior on $H_0$.  This median FoM already surpasses that of 
ConstitutionT.   However, the top-hat prior leads to significantly lower FoM.  
Under the top-hat prior we used, as many as 128 simulated datapoints are needed 
to reaches the FoM level of ConstitutionT.

\begin{figure}
    \includegraphics[width=.45\textwidth]{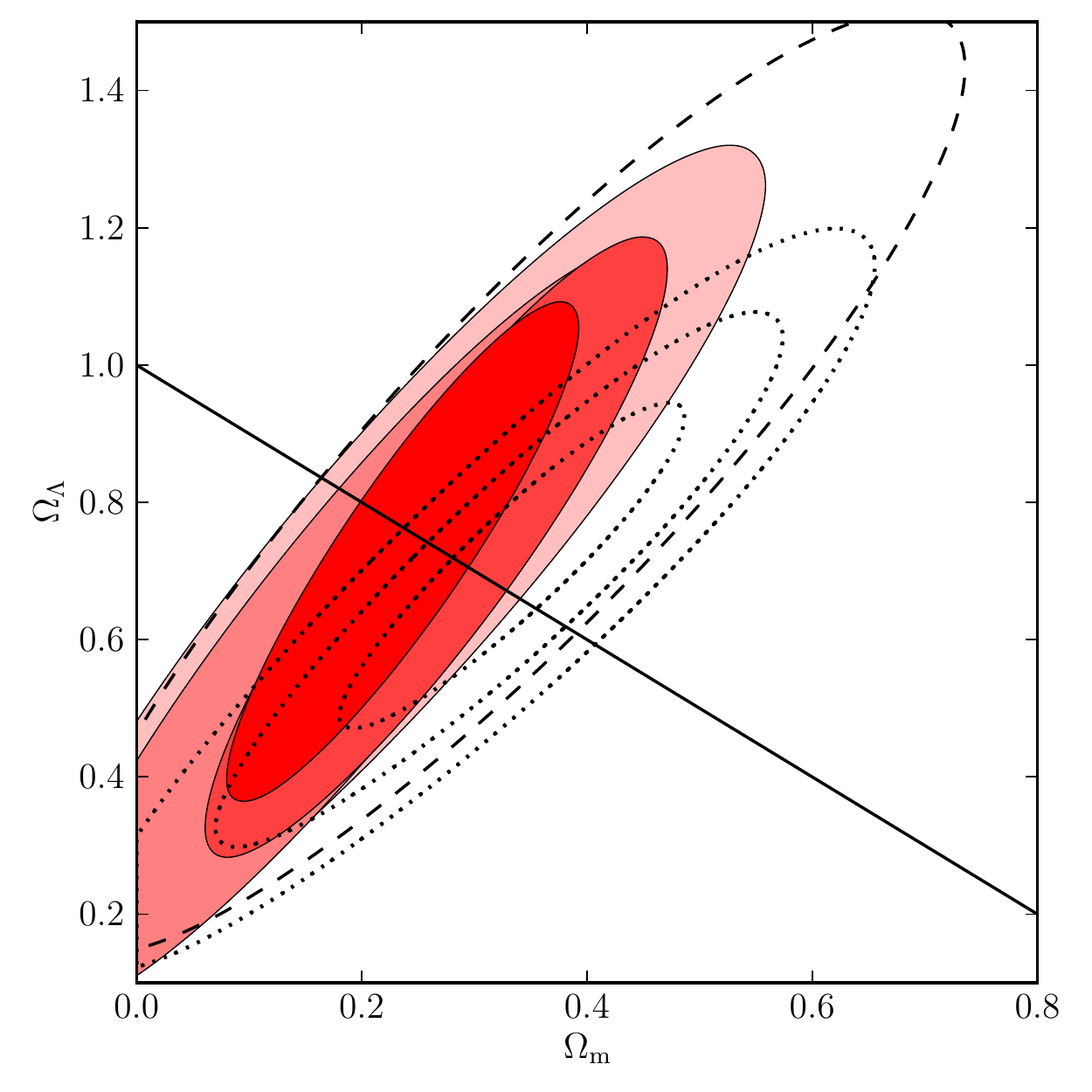}
    \caption{Confidence regions in the $(\densm, \densl)$ parameter subspace 
    calculated using the snapshot realization shown in Figure 
    \ref{fig:snapshot}.  The Gaussian prior $H_0 = 74.2 \pm 3.6\ \hunit$ is 
    assumed.  The shaded regions, from lighter to darker, correspond to the 
    $2\sigma$ constraints obtained from the data subsets containing 16, 32, 64, 
    and 128 simulated $H(z)$ datapoints respectively.  For comparison, the 
    $2\sigma$ region from pure OHD alone, with the same Gaussian prior on 
    $H_0$, is plotted in the dashed contour.  The three dotted contours are the 
    $1, 2, \text{and } 3\sigma$ constraints from the ConstitutionT SNIa 
    dataset.  The solid straight line signifies the boundary of $\densk = 0$.}
    \label{fig:fom_area}
\end{figure}

The degeneracy of confidence regions obtained from $H(z)$ data in the $(\densm, 
\densl)$ parameter subspace is shown in Figure \ref{fig:fom_area}.  Because of 
this degeneracy, $H(z)$ datasets cannot be used alone to constrain $\densk$ 
well.  This degenerate behavior is similar to that of SNIa data.

\section{Fisher Matrix Analysis of Future Data}\label{sec:future}

The simulated data used in Section \ref{sec:constraint} are based on the 
quality of currently available measurements, and we have tried not to be too 
optimistic about their uncertainties.  However, we have reasons to expect an 
increase in the quality of future $H(z)$ data.  First, 
\citet{2010MNRAS.406.2569C} analysed the observational requirement of measuring 
$H(z)$ to $3\%$ at intermediate redshifts with age-dating.  Second, the Baryon 
Oscillation Spectroscopic Survey 
(BOSS\footnote{\url{http://www.sdss3.org/boss.php}}) is designed to constrain 
$H(z)$ with $2\%$ precision at redshifts $z \approx 0.3$ and $0.6$ by measuring 
BAO imprints in the galaxy field, and at $z \approx 2.5$ using the 
Lyman-$\alpha$ absorption spectra of quasars.

By incorporating these specifications of future data, we can estimate their 
expected figures of merit using the Fisher matrix forecast technique 
\citep{1997PhRvD..56.3207D}.  The $3\times3$ Fisher matrix $\boldsymbol{F}$ is 
calculated from equation (\ref{eq:chisq}) with the $\sigma_{i}$'s determined by 
future data specifications, and the evaluation of matrix elements is made at 
the fiducial parameter values (see Section \ref{ssec:gensim}).  The matrix 
elements are the second partial derivatives of $\chi^2$ with respect of the 
parameters:
\begin{equation}
    \label{eq:fisherelems}
    F_{ij} = \frac{1}{2} \frac{\partial^2 \chi^2}{\partial \theta_i \partial 
    \theta_j},
\end{equation}
where the $\theta_i$'s are the parameters, namely $(\densm, \densl, H_0)$.  
Notice that we use the word ``Fisher matrix'' only loosely.  The second 
derivative matrix defined by equation (\ref{eq:fisherelems}) is {\em not} the 
Fisher matrix in the strict sense, but the ideas are intimately related 
\citep[see][]{2003moco.book.....D}.

In order to obtain the FoM in the 2-dimensional parameter space of $(\densm, 
\densl)$, we must marginalize over $H_0$.  We adopt the Gaussian prior on $H_0$ 
with $\sigma_H = 3.6\ \hunit$, which is the same as the one used in Section 
\ref{ssec:tmodel}.  Gaussian marginalization is performed using a 
straightforward modification of the projection technique from 
\citep{2003moco.book.....D} and \citep{2007nr..book.....P}.  A more general and 
detailed analysis of this problem was made by \citet{2010MNRAS.408..865T}, but 
for our purpose, simply adding $1 / \sigma_H^2$ to $F_{33}$ and then projecting 
onto the first two dimensions will do the work.  This is applicable when the 
mean of the prior $H_0$ is close to the fiducial value, and we have numerically 
verified the validity of this approximation in our work.

Denoting the marginalized Fisher matrix by $\widetilde{\boldsymbol{F}}$, the 
iso-$\Delta \chi^2$ contour in the parameter space is approximated by the 
quadratic equation
\begin{equation}
    \label{eq:contourofchisq}
    (\Delta \boldsymbol{\theta})^{T} \widetilde{\boldsymbol{F}} \Delta 
    \boldsymbol{\theta} = \Delta \chi^2
\end{equation}
where $\Delta \boldsymbol{\theta} = \boldsymbol{\theta} - 
\boldsymbol{\theta}_{\mathrm{fid}}$ is the parameters' deviation from the 
fiducial value, and $\Delta \chi^2$ is our chosen constant $6.17$ in Section 
\ref{ssec:fom}.  By construction, $\widetilde{\boldsymbol{F}}$ is 
positive-definite\footnote{This can be proved using the positive-definiteness 
conditions of the Schur complement 
\citep{springerlink:10.1007/0-387-24273-2_7}.}, therefore the above equation 
describes an ellipse.  Its enclosed area is simply $A = \pi / 
\sqrt{\det(\widetilde{\boldsymbol{F}} / \Delta \chi^2)}$, therefore we can 
estimate the FoM by
\begin{equation*}
    \mathrm{FoM} = \frac{1}{A} = \frac{1}{\pi} \sqrt{\det\left(\frac{1}{\Delta 
    \chi^2} \widetilde{\boldsymbol{F}} \right)}.
\end{equation*}

In our setup, we assume that the relative error of $H(z)$ anticipated by 
\citet{2010MNRAS.406.2569C} could be globally achieved within the redshift 
interval $0.1 \leq z \leq 1.0$.  Under this assumption, we can estimate how 
many datapoints will be needed to reach the ConstitutionT FoM using the Fisher 
matrix method discussed above.  We chose not to incorporate the available data 
so that we can work with the simple error model specified by future data.

\begin{figure}
    \includegraphics[width=.45\textwidth]{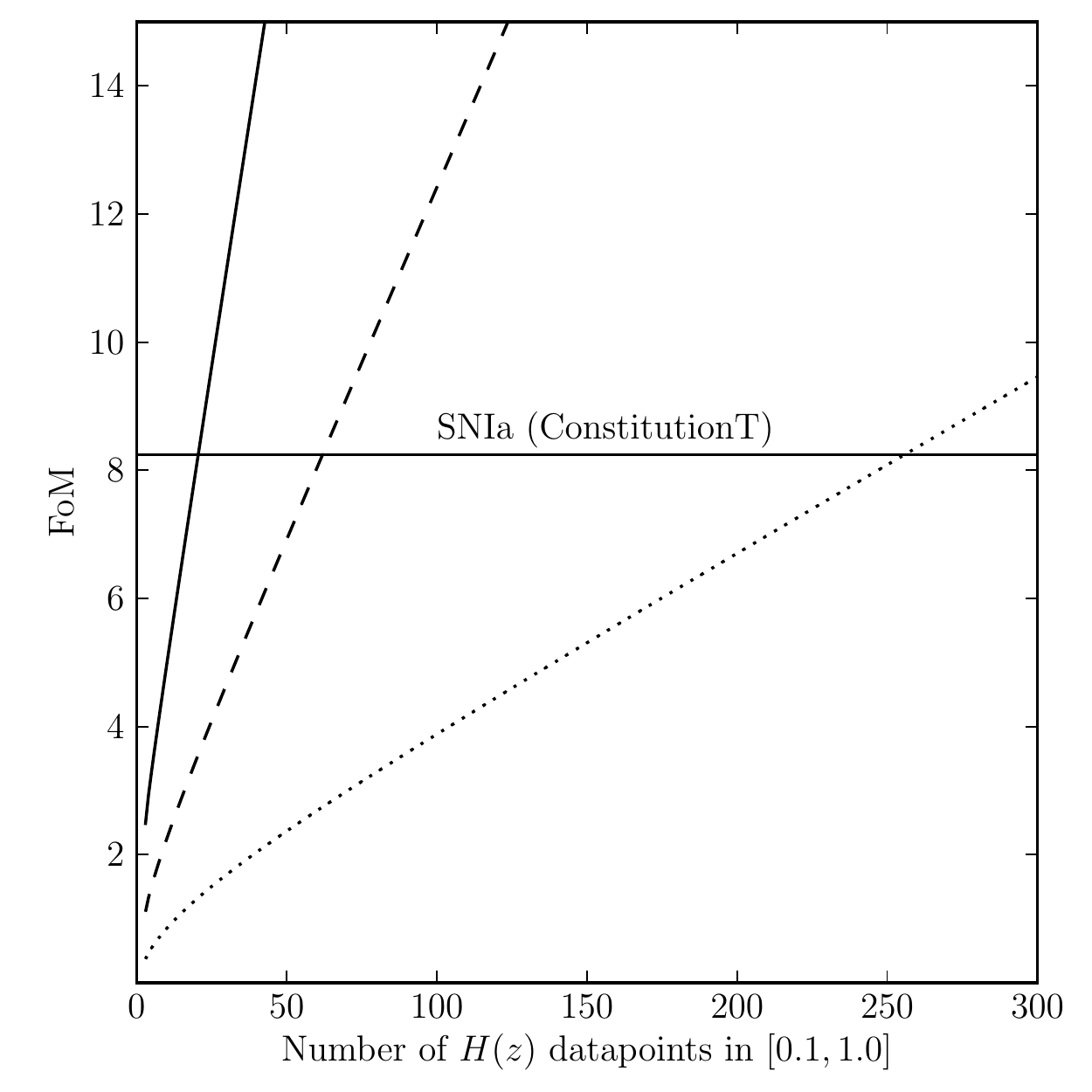}
    \caption{Predicted FoM using the error model of 
    \citep{2010MNRAS.406.2569C}.  The dependence of FoM on the number of 
    measurements is shown for three cases: the solid, dashed, and dotted lines 
    are for $3\%, 5\%,$ and $10\%$ relative error respectively.  The horizontal 
    line across the figure shows the FoM of ConsitutionT for comparison.}
    \label{fig:futurefom}
\end{figure}

\begin{figure}
    \includegraphics[width=.45\textwidth]{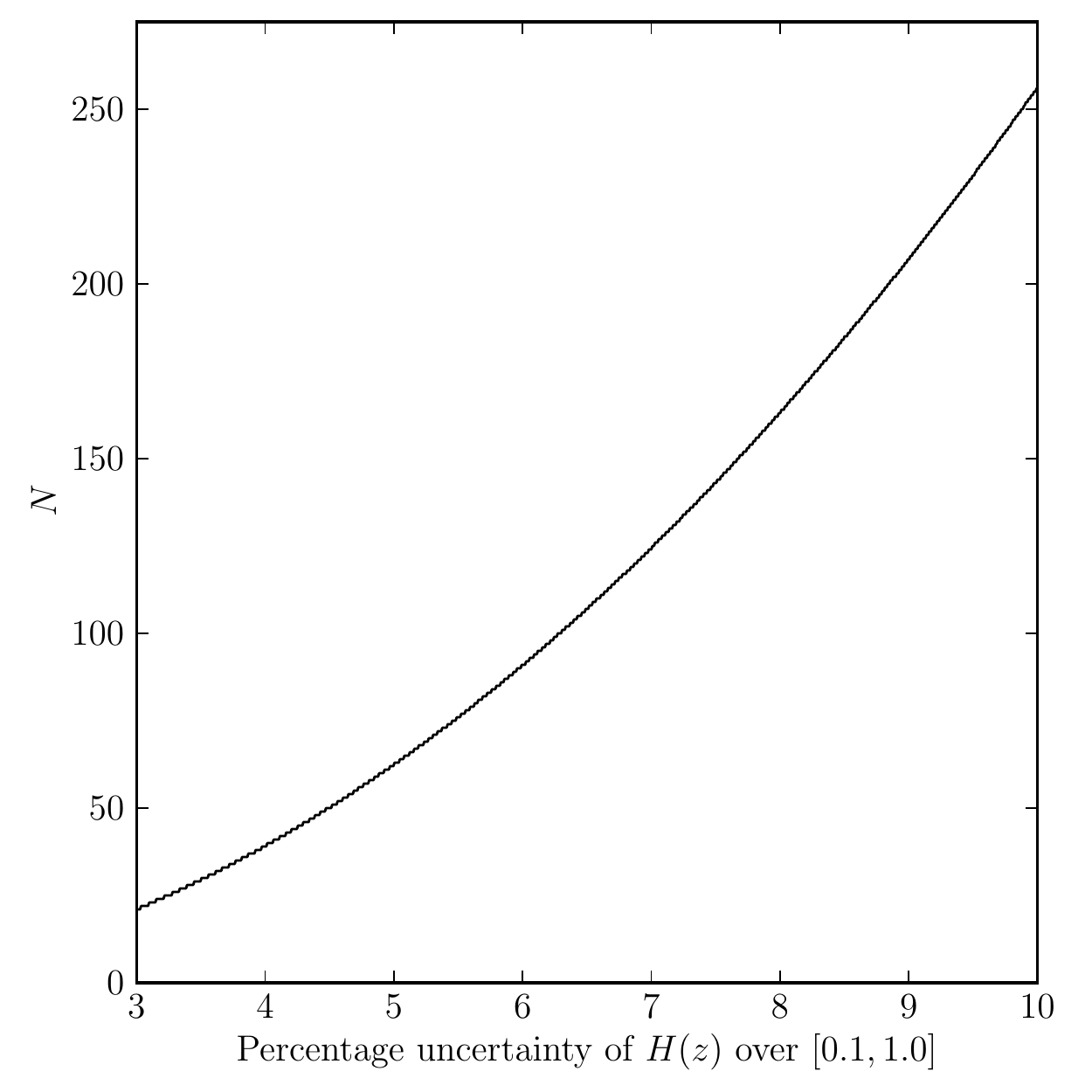}
    \caption{Number of measurements required to match ConstitutionT, $N$, as a 
    function of the relative error of $H(z)$.  The curve turns upward steeply 
    when the error is large.}
    \label{fig:requirednumber}
\end{figure}

Our main results are shown in Figures \ref{fig:futurefom} and 
\ref{fig:requirednumber}.  For relative $H(z)$ errors of $3\%, 5\%,$ and 
$10\%$, the required number of measurements are $21, 62,$ and $256$ 
respectively.  This number $N$ increases steeply as the relative error 
increases, as shown in Figure \ref{fig:requirednumber}.

\begin{figure}
    \includegraphics[width=0.45\textwidth]{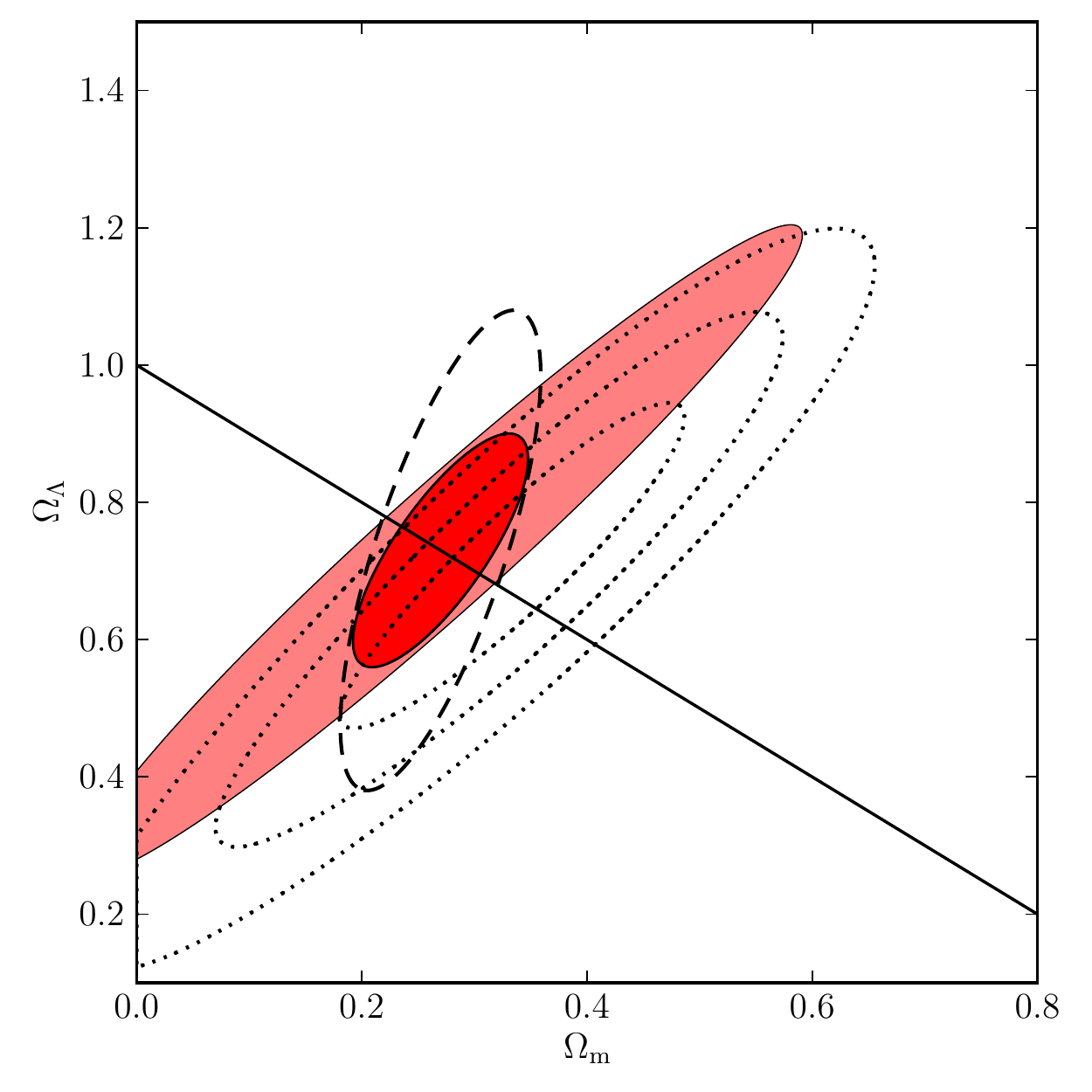}
    \caption{Same as Figure \ref{fig:fom}, but for the forecast of future
    data.  The light shaded region is the $2\sigma$ region from the 21 
    age-dating measurements with relative error $3\%$.  The unfilled region 
    bounded by the dashed contour is the $2\sigma$ region from BOSS $H(z)$.  
    The joint $2\sigma$ region of BOSS and age-dating is shown as the inner and 
    darker shaded area.  Like in Figure \ref{fig:fom}, 1, 2, and $3\sigma$ 
    constraints from ConstitutionT are shown as dotted contours for 
    comparison.}
    \label{fig:bossfom}
\end{figure}

We can also apply the Fisher matrix analysis to future BOSS data.  We find BOSS 
alone gives $\mathrm{FoM} \approx 15$.  This promising result is a combined 
consequence of its high precision and extended redshift coverage.  In Figure 
\ref{fig:bossfom} we plot the Fisher matrix forecast of the confidence regions.

\section{Conclusion and Discussions}\label{sec:conclu}

We have explored the possibility of using OHD as an alternative to SNIa 
redshift-distance data in the sense of offering comparable or higher FoM.  By 
using simulated $H(z)$ datasets with an empirical error model similar to that 
of current age-dating data, we show that more than 60 future measurements of 
$H(z)$ in the redshift range $0 \le z \le 2$ could be needed to acquire 
parameter constraints comparable with those obtained from SNIa datasets like 
ConstitutionT.  In addition, precise and accurate determination of $H_0$ is 
crucial for improving the FoM obtained from OHD, and a broad prior on $H_0$ 
leads to lower FoM.  When we progressively lower the error of future 
measurements to $3\%$ as discussed by \citep{2010MNRAS.406.2569C}, we estimate 
that $\sim$60 measurements in the shorter redshift interval $[0.1, 1.0]$ will 
be needed to achieve the same result.  In summary, we give an affirmative 
answer to the first of the two questions raised in Section \ref{sec:intro} and 
a semi-quantitative answer to the second.

Our result furthers a conclusion of \citet{2009MPLA...24.1699L} and 
\citet{2008JCAP...09..008C}, namely, that OHD play almost the same role as that 
of SNIa for the joint constraints on the $\lcdm$ model.  We have shown that the 
OHD set alone is potentially capable to be used in place of current SNIa 
datasets if it is large enough.  We note that our forecast of OHD data 
requirement is competitive in the sense of observational cost compared with 
supernova observations.  Throughout our analysis, we used the ConstitutionT 
dataset used as a standard of FoM.  This dataset is a subset of the 
Constitution compilation, a combination of the ground-based CfA3 supernova 
observations \citep{2009ApJ...700..331H} and Union, a larger compilation of 
legacy supernovae and space-based observations \citep{2008ApJ...686..749K}.  
The CfA3 sample alone requires 10 nights for each of the 185 supernovae 
observed.  On the other hand, current OHD from age-dating does not require 
space-based observations.  As described in detail by 
\citet{2010ApJS..188..280S}, 24 galaxy cluster containing target chronometer 
galaxies were obtained in only two nights using the Keck I telescope.  In 
\citep{2010MNRAS.406.2569C}, it was estimated that the South African Large 
Telescope (SALT) is capable to measure $H(z)$ to $3\%$ at an individual 
redshift in $\sim$180 hours.

Admittedly, our approach to the problem is tentative as well as 
model-dependent.  The conclusion is reached under the assumption of the 
fiducial $\lcdm$ model (eq.~[\ref{eq:fidlcdm}]), the model of uncertainties 
described in Sections \ref{ssec:gensim} and \ref{sec:future}, as well as the 
evaluation model of Section \ref{ssec:tmodel}.  Therefore, care must be taken 
not to extrapolate our conclusions well beyond those assumptions, especially 
when used in the planning of observations.

Even if our uncertainty models fit real-world observations well, in practice 
the results may be less promising than what we suggested in this paper.  We 
assumed a fairly deep redshift coverage which may be difficult to reach by 
current observations \citep[see][for the redshift distribution of LRGs used to 
deduce SJVKS10]{2010ApJS..188..280S}.  Ultimately, the FoM from OHD may be 
limited by our ability to select enough samples for future observation and the 
redshift distribution of these samples, rather than by the relative error of 
each individual measurement.  Fortunately, the BOSS project could extend $H(z)$ 
measurement into deeper redshift.  Also, It is worth noting that the proposed 
Sandage-Loeb observational plan 
\citep{2007PhRvD..75f2001C,2010PhLB..691...11Z,2010PhRvD..82l3513A} could be 
used to extend our knowledge of cosmic expansion into the even deeper redshift 
realm of $2 \le z \le 5$ by measuring the secular variation of cosmological 
redshifts of the Lyman-$\alpha$ forest with future high-resolution, space-based 
spectroscopic instruments (see \citep{1962ApJ...136..319S} and 
\citep{1998ApJ...499L.111L} for the foundations of the test).

Finally, we note that future CMB observation programs, such as the Atacama 
Cosmology 
Telescope\footnote{\url{http://www.physics.princeton.edu/act/index.html}}, may 
be able to identify more than 2000 passively evolving galaxies up to $z \approx 
1.5$ via the Sunyaev-Zel'dovich effect, and their spectra can be analyzed to 
yield age measurements that will yield approximately 1000 $H(z)$ determinations 
with $15\%$ error \citep{2005PhRvD..71l3001S}.  This promises a future data 
capacity an order of magnitude more than what we have estimated to be enough to 
match current SNIa datasets.  Combining this prospect with future high-$z$, 
high-accuracy $H(z)$ determinations from BAO observations, it is reasonable to 
expect that the OHD will play an increasingly important role in the future 
study of the expansion history of the universe and cosmological parameters.

\acknowledgements

We thank the anonymous referee whose suggestions greatly helped us improve this 
paper.  CM is grateful to Chen-Tao Yang for useful discussions.  This work was 
supported by the National Science Foundation of China (Grants No.~10473002), 
the Ministry of Science and Technology National Basic Science program (project 
973) under grant No.~2009CB24901, the Fundamental Research Funds for the 
Central Universities.

\bibliographystyle{hapj}
\bibliography{hzrev}

\begin{thebibliography}{54}
\expandafter\ifx\csname natexlab\endcsname\relax\def\natexlab#1{#1}\fi

\bibitem[{{Albrecht} {et~al.}(2006){Albrecht}, {Bernstein}, {Cahn}, {Freedman},
  {Hewitt}, {Hu}, {Huth}, {Kamionkowski}, {Kolb}, {Knox}, {Mather}, {Staggs},
  \& {Suntzeff}}]{2006astro.ph..9591A}
{Albrecht}, A. {et~al.} 2006, preprint, arXiv:astro-ph/0609591

\bibitem[{{Aldering} {et~al.}(2004){Aldering}, {Althouse}, {Amanullah},
  {Annis}, {Astier}, {Baltay}, {Barrelet}, {Basa}, {Bebek}, {Bergstrom},
  {Bernstein}, {Bester}, {Bigelow}, {Blandford}, {Bohlin}, {Bonissent},
  {Bower}, {Brown}, {Campbell}, {Carithers}, {Commins}, {Craig}, {Day},
  {DeJongh}, {Deustua}, {Diehl}, {Dodelson}, {Ealet}, {Ellis}, {Emmet},
  {Fouchez}, {Frieman}, {Fruchter}, {Gerdes}, {Gladney}, {Goldhaber}, {Goobar},
  {Groom}, {Heetderks}, {Hoff}, {Holland}, {Huffer}, {Hui}, {Huterer}, {Jain},
  {Jelinsky}, {Karcher}, {Kent}, {Kahn}, {Kim}, {Kolbe}, {Krieger}, {Kushner},
  {Kuznetsova}, {Lafever}, {Lamoureux}, {Lampton}, {Le Fevre}, {Levi}, {Limon},
  {Lin}, {Linder}, {Loken}, {Lorenzon}, {Malina}, {Marriner}, {Marshall},
  {Massey}, {Mazure}, {McKay}, {McKee}, {Miquel}, {Morgan}, {Mortsell},
  {Mostek}, {Mufson}, {Musser}, {Nugent}, {Oluseyi}, {Pain}, {Palaio},
  {Pankow}, {Peoples}, {Perlmutter}, {Prieto}, {Rabinowitz}, {Refregier},
  {Rhodes}, {Roe}, {Rusin}, {Scarpine}, {Schubnell}, {Sholl}, {Smadja},
  {Smith}, {Smoot}, {Snyder}, {Spadafora}, {Stebbins}, {Stoughton},
  {Szymkowiak}, {Tarle}, {Taylor}, {Tilquin}, {Tomasch}, {Tucker}, {Vincent},
  {von der Lippe}, {Walder}, {Wang}, \& {Wester}}]{2004astro.ph..5232S}
{Aldering}, G. {et~al.} 2004, preprint, arXiv:astro-ph/0405232

\bibitem[{{Ara{\'u}jo} \& {Stoeger}(2010)}]{2010PhRvD..82l3513A}
{Ara{\'u}jo}, M.~E., \& {Stoeger}, W.~R. 2010, \prd, 82, 123513,
  arXiv:1009.2783

\bibitem[{{Blake} \& {Glazebrook}(2003)}]{2003ApJ...594..665B}
{Blake}, C., \& {Glazebrook}, K. 2003, \apj, 594, 665, arXiv:astro-ph/0301632

\bibitem[{{Bueno Sanchez} {et~al.}(2009){Bueno Sanchez}, {Nesseris}, \&
  {Perivolaropoulos}}]{2009JCAP...11..029B}
{Bueno Sanchez}, J.~C., {Nesseris}, S., \& {Perivolaropoulos}, L. 2009,
  J.~Cosmol.~Astropart.~Phys., 11, 29, arXiv:0908.2636

\bibitem[{{Carvalho} {et~al.}(2008){Carvalho}, {Santos}, {Alcaniz}, \&
  {Santos}}]{2008JCAP...09..008C}
{Carvalho}, F.~C., {Santos}, E.~M., {Alcaniz}, J.~S., \& {Santos}, J. 2008,
  J.~Cosmol.~Astropart.~Phys., 9, 8, arXiv:0804.2878

\bibitem[{{Chen} {et~al.}(2003){Chen}, {Gott III}, \&
  {Ratra}}]{2003PASP..115.1269C}
{Chen}, G., {Gott III}, J.~R., \& {Ratra}, B. 2003, \pasp, 115, 1269,
  arXiv:astro-ph/0308099

\bibitem[{{Corasaniti} {et~al.}(2007){Corasaniti}, {Huterer}, \&
  {Melchiorri}}]{2007PhRvD..75f2001C}
{Corasaniti}, P., {Huterer}, D., \& {Melchiorri}, A. 2007, \prd, 75, 062001,
  arXiv:astro-ph/0701433

\bibitem[{{Crawford} {et~al.}(2010){Crawford}, {Ratsimbazafy}, {Cress},
  {Olivier}, {Blyth}, \& {van der Heyden}}]{2010MNRAS.406.2569C}
{Crawford}, S.~M., {Ratsimbazafy}, A.~L., {Cress}, C.~M., {Olivier}, E.~A.,
  {Blyth}, S., \& {van der Heyden}, K.~J. 2010, \mnras, 406, 2569,
  arXiv:1004.2378

\bibitem[{{Dodelson}(2003)}]{2003moco.book.....D}
{Dodelson}, S. 2003, {Modern Cosmology} ({San Diego, CA}: {Academic Press})

\bibitem[{{Dodelson} {et~al.}(1997){Dodelson}, {Kinney}, \&
  {Kolb}}]{1997PhRvD..56.3207D}
{Dodelson}, S., {Kinney}, W.~H., \& {Kolb}, E.~W. 1997, \prd, 56, 3207,
  arXiv:astro-ph/9702166

\bibitem[{{Eisenstein} {et~al.}(2005){Eisenstein}, {Zehavi}, {Hogg},
  {Scoccimarro}, {Blanton}, {Nichol}, {Scranton}, {Seo}, {Tegmark}, {Zheng},
  {Anderson}, {Annis}, {Bahcall}, {Brinkmann}, {Burles}, {Castander},
  {Connolly}, {Csabai}, {Doi}, {Fukugita}, {Frieman}, {Glazebrook}, {Gunn},
  {Hendry}, {Hennessy}, {Ivezi{\'c}}, {Kent}, {Knapp}, {Lin}, {Loh}, {Lupton},
  {Margon}, {McKay}, {Meiksin}, {Munn}, {Pope}, {Richmond}, {Schlegel},
  {Schneider}, {Shimasaku}, {Stoughton}, {Strauss}, {SubbaRao}, {Szalay},
  {Szapudi}, {Tucker}, {Yanny}, \& {York}}]{2005ApJ...633..560E}
{Eisenstein}, D.~J. {et~al.} 2005, \apj, 633, 560, arXiv:astro-ph/0501171

\bibitem[{{Freedman} {et~al.}(2001){Freedman}, {Madore}, {Gibson}, {Ferrarese},
  {Kelson}, {Sakai}, {Mould}, {Kennicutt}, {Ford}, {Graham}, {Huchra},
  {Hughes}, {Illingworth}, {Macri}, \& {Stetson}}]{2001ApJ...553...47F}
{Freedman}, W.~L. {et~al.} 2001, \apj, 553, 47, arXiv:astro-ph/0012376

\bibitem[{{Gazta{\~n}aga} {et~al.}(2009){Gazta{\~n}aga}, {Cabr{\'e}}, \&
  {Hui}}]{2009MNRAS.399.1663G}
{Gazta{\~n}aga}, E., {Cabr{\'e}}, A., \& {Hui}, L. 2009, \mnras, 399, 1663,
  arXiv:0807.3551

\bibitem[{{Ghirlanda} {et~al.}(2004){Ghirlanda}, {Ghisellini}, {Lazzati}, \&
  {Firmani}}]{2004ApJ...613L..13G}
{Ghirlanda}, G., {Ghisellini}, G., {Lazzati}, D., \& {Firmani}, C. 2004, \apjl,
  613, L13, arXiv:astro-ph/0408350

\bibitem[{{Gott III} {et~al.}(2001){Gott III}, {Vogeley}, {Podariu}, \&
  {Ratra}}]{2001ApJ...549....1G}
{Gott III}, J.~R., {Vogeley}, M.~S., {Podariu}, S., \& {Ratra}, B. 2001, \apj,
  549, 1, arXiv:astro-ph/0006103

\bibitem[{{Hicken} {et~al.}(2009{\natexlab{a}}){Hicken}, {Challis}, {Jha},
  {Kirshner}, {Matheson}, {Modjaz}, {Rest}, {Wood-Vasey}, {Bakos}, {Barton},
  {Berlind}, {Bragg}, {Brice{\~n}o}, {Brown}, {Caldwell}, {Calkins}, {Cho},
  {Ciupik}, {Contreras}, {Dendy}, {Dosaj}, {Durham}, {Eriksen}, {Esquerdo},
  {Everett}, {Falco}, {Fernandez}, {Gaba}, {Garnavich}, {Graves}, {Green},
  {Groner}, {Hergenrother}, {Holman}, {Hradecky}, {Huchra}, {Hutchison},
  {Jerius}, {Jordan}, {Kilgard}, {Krauss}, {Luhman}, {Macri}, {Marrone},
  {McDowell}, {McIntosh}, {McNamara}, {Megeath}, {Mochejska}, {Munoz},
  {Muzerolle}, {Naranjo}, {Narayan}, {Pahre}, {Peters}, {Peterson}, {Rines},
  {Ripman}, {Roussanova}, {Schild}, {Sicilia-Aguilar}, {Sokoloski}, {Smalley},
  {Smith}, {Spahr}, {Stanek}, {Barmby}, {Blondin}, {Stubbs}, {Szentgyorgyi},
  {Torres}, {Vaz}, {Vikhlinin}, {Wang}, {Westover}, {Woods}, \&
  {Zhao}}]{2009ApJ...700..331H}
{Hicken}, M. {et~al.} 2009{\natexlab{a}}, \apj, 700, 331, arXiv:0901.4787

\bibitem[{{Hicken} {et~al.}(2009{\natexlab{b}}){Hicken}, {Wood-Vasey},
  {Blondin}, {Challis}, {Jha}, {Kelly}, {Rest}, \&
  {Kirshner}}]{2009ApJ...700.1097H}
{Hicken}, M., {Wood-Vasey}, W.~M., {Blondin}, S., {Challis}, P., {Jha}, S.,
  {Kelly}, P.~L., {Rest}, A., \& {Kirshner}, R.~P. 2009{\natexlab{b}}, \apj,
  700, 1097, arXiv:0901.4804

\bibitem[{{Huterer}(2009)}]{2009NuPhS.194..239H}
{Huterer}, D. 2009, Nuclear Physics B Proceedings Supplements, 194, 239

\bibitem[{{Jimenez} \& {Loeb}(2002)}]{2002ApJ...573...37J}
{Jimenez}, R., \& {Loeb}, A. 2002, \apj, 573, 37, arXiv:astro-ph/0106145

\bibitem[{{Jimenez} {et~al.}(2003){Jimenez}, {Verde}, {Treu}, \&
  {Stern}}]{2003ApJ...593..622J}
{Jimenez}, R., {Verde}, L., {Treu}, T., \& {Stern}, D. 2003, \apj, 593, 622,
  arXiv:astro-ph/0302560

\bibitem[{{Kim} {et~al.}(2004){Kim}, {Linder}, {Miquel}, \&
  {Mostek}}]{2004MNRAS.347..909K}
{Kim}, A.~G., {Linder}, E.~V., {Miquel}, R., \& {Mostek}, N. 2004, \mnras, 347,
  909, arXiv:astro-ph/0304509

\bibitem[{{Komatsu} {et~al.}(2010){Komatsu}, {Smith}, {Dunkley}, {Bennett},
  {Gold}, {Hinshaw}, {Jarosik}, {Larson}, {Nolta}, {Page}, {Spergel},
  {Halpern}, {Hill}, {Kogut}, {Limon}, {Meyer}, {Odegard}, {Tucker}, {Weiland},
  {Wollack}, \& {Wright}}]{2010arXiv1001.4538K}
{Komatsu}, E. {et~al.} 2010, \apjs, submitted, arXiv:1001.4538

\bibitem[{{Kowalski} {et~al.}(2008){Kowalski}, {Rubin}, {Aldering},
  {Agostinho}, {Amadon}, {Amanullah}, {Balland}, {Barbary}, {Blanc}, {Challis},
  {Conley}, {Connolly}, {Covarrubias}, {Dawson}, {Deustua}, {Ellis}, {Fabbro},
  {Fadeyev}, {Fan}, {Farris}, {Folatelli}, {Frye}, {Garavini}, {Gates},
  {Germany}, {Goldhaber}, {Goldman}, {Goobar}, {Groom}, {Haissinski}, {Hardin},
  {Hook}, {Kent}, {Kim}, {Knop}, {Lidman}, {Linder}, {Mendez}, {Meyers},
  {Miller}, {Moniez}, {Mour{\~a}o}, {Newberg}, {Nobili}, {Nugent}, {Pain},
  {Perdereau}, {Perlmutter}, {Phillips}, {Prasad}, {Quimby}, {Regnault},
  {Rich}, {Rubenstein}, {Ruiz-Lapuente}, {Santos}, {Schaefer}, {Schommer},
  {Smith}, {Soderberg}, {Spadafora}, {Strolger}, {Strovink}, {Suntzeff},
  {Suzuki}, {Thomas}, {Walton}, {Wang}, {Wood-Vasey}, \&
  {Yun}}]{2008ApJ...686..749K}
{Kowalski}, M. {et~al.} 2008, \apj, 686, 749, arXiv:0804.4142

\bibitem[{{Li} {et~al.}(2008){Li}, {Xia}, {Liu}, {Zhao}, {Fan}, \&
  {Zhang}}]{2008ApJ...680...92L}
{Li}, H., {Xia}, J.-Q., {Liu}, J., {Zhao}, G.-B., {Fan}, Z.-H., \& {Zhang}, X.
  2008, \apj, 680, 92, arXiv:0711.1792

\bibitem[{{Li} {et~al.}(2009){Li}, {Li}, {Wang}, \&
  {Zhang}}]{2009JCAP...06..036L}
{Li}, M., {Li}, X.-D., {Wang}, S., \& {Zhang}, X. 2009,
  J.~Cosmol.~Astropart.~Phys., 6, 36, arXiv:0904.0928

\bibitem[{{Liang} {et~al.}(2010){Liang}, {Wu}, \&
  {Zhang}}]{2010PhRvD..81h3518L}
{Liang}, N., {Wu}, P., \& {Zhang}, S.~N. 2010, \prd, 81, 083518,
  arXiv:0911.5644

\bibitem[{{Lin} {et~al.}(2009){Lin}, {Hao}, {Wang}, {Yuan}, {Yi}, {Zhang}, \&
  {Wang}}]{2009MPLA...24.1699L}
{Lin}, H., {Hao}, C., {Wang}, X., {Yuan}, Q., {Yi}, Z.-L., {Zhang}, T.-J., \&
  {Wang}, B.-Q. 2009, Mod. Phys. Lett. A, 24, 1699, arXiv:0804.3135

\bibitem[{{Linder}(2006)}]{2006APh....26..102L}
{Linder}, E.~V. 2006, Astropart.~Phys., 26, 102, arXiv:astro-ph/0604280

\bibitem[{{Liu} {et~al.}(2008){Liu}, {Li}, {Hao}, \&
  {Jin}}]{2008MNRAS.388..275L}
{Liu}, D.-J., {Li}, X.-Z., {Hao}, J., \& {Jin}, X.-H. 2008, \mnras, 388, 275,
  arXiv:0804.3829

\bibitem[{{Loeb}(1998)}]{1998ApJ...499L.111L}
{Loeb}, A. 1998, \apjl, 499, L111, arXiv:astro-ph/9802112

\bibitem[{{Mignone} \& {Bartelmann}(2008)}]{2008A&A...481..295M}
{Mignone}, C., \& {Bartelmann}, M. 2008, \aap, 481, 295, arXiv:0711.0370

\bibitem[{NIST(2003)}]{nistsematech}
NIST. 2003, {NIST/SEMATECH e-Handbook of Statistical Methods}
  ({Gaithersburg,~MD}: National Institute of Standards and Technology),
  \url{http://www.itl.nist.gov/div898/handbook/index.htm}

\bibitem[{{Percival} {et~al.}(2010){Percival}, {Reid}, {Eisenstein}, {Bahcall},
  {Budavari}, {Frieman}, {Fukugita}, {Gunn}, {Ivezi{\'c}}, {Knapp}, {Kron},
  {Loveday}, {Lupton}, {McKay}, {Meiksin}, {Nichol}, {Pope}, {Schlegel},
  {Schneider}, {Spergel}, {Stoughton}, {Strauss}, {Szalay}, {Tegmark},
  {Vogeley}, {Weinberg}, {York}, \& {Zehavi}}]{2010MNRAS.401.2148P}
{Percival}, W.~J. {et~al.} 2010, \mnras, 401, 2148, arXiv:0907.1660

\bibitem[{{Perlmutter} {et~al.}(1999){Perlmutter}, {Aldering}, {Goldhaber},
  {Knop}, {Nugent}, {Castro}, {Deustua}, {Fabbro}, {Goobar}, {Groom}, {Hook},
  {Kim}, {Kim}, {Lee}, {Nunes}, {Pain}, {Pennypacker}, {Quimby}, {Lidman},
  {Ellis}, {Irwin}, {McMahon}, {Ruiz-Lapuente}, {Walton}, {Schaefer}, {Boyle},
  {Filippenko}, {Matheson}, {Fruchter}, {Panagia}, {Newberg}, {Couch}, \& {The
  Supernova Cosmology Project}}]{1999ApJ...517..565P}
{Perlmutter}, S. {et~al.} 1999, \apj, 517, 565, arXiv:astro-ph/9812133

\bibitem[{{Press} {et~al.}(2007){Press}, {Teukolsky}, {Vetterling}, \&
  {Flannery}}]{2007nr..book.....P}
{Press}, W.~H., {Teukolsky}, S.~A., {Vetterling}, W.~T., \& {Flannery}, B.~P.
  2007, {Numerical Recipes: the Art of Scientific Computing}, 3rd edn.
  ({Cambridge}: {Cambridge Univ.~Press})

\bibitem[{{Puntanen} \& {Styan}(2005)}]{springerlink:10.1007/0-387-24273-2_7}
{Puntanen}, S., \& {Styan}, G.~P.~H. 2005, in {Numerical Methods and
  Algorithms}, Vol.~4, The Schur Complement and Its Applications, ed.
  C.~{Brezinski} \& F.~{Zhang} ({New York}: Springer), 163--226

\bibitem[{{Riess} {et~al.}(1998){Riess}, {Filippenko}, {Challis},
  {Clocchiatti}, {Diercks}, {Garnavich}, {Gilliland}, {Hogan}, {Jha},
  {Kirshner}, {Leibundgut}, {Phillips}, {Reiss}, {Schmidt}, {Schommer},
  {Smith}, {Spyromilio}, {Stubbs}, {Suntzeff}, \&
  {Tonry}}]{1998AJ....116.1009R}
{Riess}, A.~G. {et~al.} 1998, \aj, 116, 1009, arXiv:astro-ph/9805201

\bibitem[{{Riess} {et~al.}(2009){Riess}, {Macri}, {Casertano}, {Sosey},
  {Lampeitl}, {Ferguson}, {Filippenko}, {Jha}, {Li}, {Chornock}, \&
  {Sarkar}}]{2009ApJ...699..539R}
------. 2009, \apj, 699, 539, arXiv:0905.0695

\bibitem[{{Samushia} \& {Ratra}(2006)}]{2006ApJ...650L...5S}
{Samushia}, L., \& {Ratra}, B. 2006, \apjl, 650, L5, arXiv:astro-ph/0607301

\bibitem[{{Sandage}(1962)}]{1962ApJ...136..319S}
{Sandage}, A. 1962, \apj, 136, 319

\bibitem[{{Seo} \& {Eisenstein}(2005)}]{2005ApJ...633..575S}
{Seo}, H.-J., \& {Eisenstein}, D.~J. 2005, \apj, 633, 575,
  arXiv:astro-ph/0507338

\bibitem[{{Shafieloo} {et~al.}(2006){Shafieloo}, {Alam}, {Sahni}, \&
  {Starobinsky}}]{2006MNRAS.366.1081S}
{Shafieloo}, A., {Alam}, U., {Sahni}, V., \& {Starobinsky}, A.~A. 2006, \mnras,
  366, 1081, arXiv:astro-ph/0505329

\bibitem[{{Simon} {et~al.}(2005){Simon}, {Verde}, \&
  {Jimenez}}]{2005PhRvD..71l3001S}
{Simon}, J., {Verde}, L., \& {Jimenez}, R. 2005, \prd, 71, 123001,
  arXiv:astro-ph/0412269

\bibitem[{{Spergel} {et~al.}(2007){Spergel}, {Bean}, {Dor{\'e}}, {Nolta},
  {Bennett}, {Dunkley}, {Hinshaw}, {Jarosik}, {Komatsu}, {Page}, {Peiris},
  {Verde}, {Halpern}, {Hill}, {Kogut}, {Limon}, {Meyer}, {Odegard}, {Tucker},
  {Weiland}, {Wollack}, \& {Wright}}]{2007ApJS..170..377S}
{Spergel}, D.~N. {et~al.} 2007, \apjs, 170, 377, arXiv:astro-ph/0603449

\bibitem[{{Stern} {et~al.}(2010{\natexlab{a}}){Stern}, {Jimenez}, {Verde},
  {Kamionkowski}, \& {Stanford}}]{2010JCAP...02..008S}
{Stern}, D., {Jimenez}, R., {Verde}, L., {Kamionkowski}, M., \& {Stanford},
  S.~A. 2010{\natexlab{a}}, J.~Cosmol.~Astropart.~Phys., 2, 8, arXiv:0907.3149

\bibitem[{{Stern} {et~al.}(2010{\natexlab{b}}){Stern}, {Jimenez}, {Verde},
  {Stanford}, \& {Kamionkowski}}]{2010ApJS..188..280S}
{Stern}, D., {Jimenez}, R., {Verde}, L., {Stanford}, S.~A., \& {Kamionkowski},
  M. 2010{\natexlab{b}}, \apjs, 188, 280, arXiv:0907.3152

\bibitem[{{Taylor} \& {Kitching}(2010)}]{2010MNRAS.408..865T}
{Taylor}, A.~N., \& {Kitching}, T.~D. 2010, \mnras, 408, 865, arXiv:1003.1136

\bibitem[{{Wang} \& {Tegmark}(2004)}]{2004PhRvL..92x1302W}
{Wang}, Y., \& {Tegmark}, M. 2004, \prl, 92, 241302, arXiv:astro-ph/0403292

\bibitem[{{Wang} \& {Tegmark}(2005)}]{2005PhRvD..71j3513W}
------. 2005, \prd, 71, 103513, arXiv:astro-ph/0501351

\bibitem[{{Wei}(2010)}]{2010PhLB..687..286W}
{Wei}, H. 2010, Phys.~Lett.~B, 687, 286, arXiv:0906.0828

\bibitem[{{Yang} \& {Chen}(2009)}]{2009MNRAS.394.1449Y}
{Yang}, X.-J., \& {Chen}, D.-M. 2009, \mnras, 394, 1449, arXiv:0812.0660

\bibitem[{{Yi} \& {Zhang}(2007)}]{2007MPLA...22...41Y}
{Yi}, Z.-L., \& {Zhang}, T.-J. 2007, Mod.~Phys.~Lett.~A, 22, 41,
  arXiv:astro-ph/0605596

\bibitem[{{Zhang} {et~al.}(2010){Zhang}, {Zhang}, \&
  {Zhang}}]{2010PhLB..691...11Z}
{Zhang}, J., {Zhang}, L., \& {Zhang}, X. 2010, Phys.~Lett.~B, 691, 11,
  arXiv:1006.1738

\end{thebibliography}

\end{document}